\documentclass[prd, aps, nofootinbib, preprintnumbers, 
 showpacs, superscriptaddress, twocolumn]{revtex4}

\usepackage{amsmath}
\usepackage{amssymb}
\usepackage[english]{babel}
\usepackage{color}
\usepackage{dcolumn}
\usepackage{graphicx}
\usepackage{epsfig}
\usepackage{epstopdf}
\usepackage[latin9]{inputenc}
\usepackage{subfigure}

\newcommand{\lessthansimilarto}{\lower3pt\hbox{$\buildrel{<}\over{\sim}$}}
\newcommand{\greaterthansimilarto}{\lower3pt\hbox{$\buildrel{>}\over{\sim}$}}

\newcommand{\be}{\begin{equation}}
\newcommand{\ee}{\end{equation}}

\sloppypar

\begin{document}

\title{Excision without excision}

\author{David~Brown}

\affiliation{Department of Physics, North Carolina State University,
  Raleigh, NC 27695, USA}

\author{Olivier~Sarbach}

\affiliation{Instituto de F\a'{\i}sica y Matem\a'aticas,
Universidad Michoacana de San Nicol\a'as de Hidalgo,
Morelia, Michoac\a'an, M\a'exico}

\author{Erik~Schnetter}

\affiliation{Center for Computation \& Technology,
  Louisiana State University, Baton Rouge, LA, USA}

\affiliation{Department of Physics and Astronomy,
  Louisiana State University, Baton Rouge, LA, USA}

\author{Manuel~Tiglio}

\affiliation{Department of Physics and Astronomy,
  Louisiana State University, Baton Rouge, LA, USA}

\affiliation{Center for Computation \& Technology,
  Louisiana State University, Baton Rouge, LA, USA}

\author{Peter~Diener}

\affiliation{Department of Physics and Astronomy,
  Louisiana State University, Baton Rouge, LA, USA}

\affiliation{Center for Computation \& Technology,
  Louisiana State University, Baton Rouge, LA, USA}

\author{Ian~Hawke}

\affiliation{School of Mathematics, University of Southampton,
  Southampton, UK}

\author{Denis~Pollney}

\affiliation{Albert-Einstein-Institut, Max-Planck-Institut f\a"ur
  Gravitationsphysik, Golm, Germany}

\date{July 19, 2007}

\begin{abstract}
\emph{to turducken (turduckens, turduckening, turduckened, turduckened)}
  [math.]: To stuff a black hole.
  \vspace{3ex}

  We analyze and apply an alternative to
  black hole excision based on 
  smoothing the interior of black holes
  with arbitrary -- possibly constraint violating -- initial data, and solving 
  the vacuum Einstein evolution equations everywhere. 
  By deriving the constraint propagation system for
  our hyperbolic formulation of the BSSN
  evolution system we rigorously prove that the constraints propagate
  causally and so any constraint violations introduced inside the
  black holes cannot affect the exterior spacetime.
  (This does not follow from the causal structure of the spacetime as
  is often assumed.)
  We present numerical evolutions of Cook-Pfeiffer binary black hole
  initial configurations showing that
  these techniques appear to work robustly for generic data.
  We also present numerical evidence from spherically symmetric evolutions
  that for the gauge conditions used the same stationary end-state
  is approached irrespective of the choice of initial data and 
  smoothing procedure. 

\end{abstract}


\maketitle

\section{Introduction}

Currently, there are essentially two different ways of dealing with 
singularities in the numerical evolution of orbiting black holes. One 
technique is black hole excision
\cite{Unruh84-note, Thornburg87}, where the interior of each black hole is removed from the
 computational domain by an inner boundary. The other is the so called ``moving
punctures'' technique \cite{Campanelli:2005dd,Baker:2005vv}, where the initial
asymptotically flat regions inside each black hole are represented by
``puncture points''.
Long, multi-orbit binary black hole
simulations have been achieved over the last few years using both excision 
\cite{Pretorius:2007pd,Pfeiffer:2007} and moving punctures 
(see \cite{Bruegmann:2007} and references therein).


The puncture technique does not make use of the black hole excision
idea, at least not in the classical sense of placing an inner boundary
inside each black hole. Instead, the fields that initially describe
the puncture points are allowed to evolve freely in the (topologically
trivial) computational domain and the subtleties of black hole excision
are replaced by the subtleties involved in approximating the
singularities in the equations at the puncture points. The particular
appeal of the ``moving punctures'' technique compared to black hole
excision is that it appears to be simpler to achieve a stable
discretization near the puncture points than at an excision boundary;
however, there appears to be an implicit limitation of the method in
that it is in principle tied to the use of puncture data. Recently,
light has been shed on the geometric picture behind moving
punctures \cite{Hannam:2006vv,Brown:2007,Garfinkle2007a}.

In this paper we discuss a technique for evolving black holes which
shares the simplicity of moving punctures but is not
restricted to puncture-type initial data and does not need any
regularization of the equations near special points. The method also relies
on the intuitive idea behind black hole excision that ``no
physical information can escape from the interior of a black hole'',
but proceeds in a different way. In particular, it does not require
placing an inner boundary per black hole in order to remove the
interiors. The computational domain in this technique is trivial (from
a topological point of view) and the discretization therefore
remains simple.

The basic idea is the following: if no physical information can leave
the interior of the black hole, why not just change the interior 
to one's advantage? The spirit of this idea is not new, and has been
advocated for a long time 
in several forms, most notably by Bona and collaborators
\cite{Arbona97, Arbona99a, Bona04a}
and by
Misner \cite{Misner01}. In particular, in \cite{Bona04a}, a ``free black
evolution'' approach was advocated, where the interior of each 
black hole is smoothed with arbitrary data and the
vacuum Einstein evolution equations are solved everywhere. In general the smoothing process 
generates constraint violations. Thus, a key ingredient of this approach is
to guarantee that the form of the equations does not allow for constraint
violations to propagate to the
outside. This is highly non-trivial. In fact, it is well known that 
depending on the form of the Einstein equations
used, gauge and constraint modes can propagate with
arbitrary (including superluminal) speeds and, in particular, constraint violations \emph{can} 
leak from the interior of black holes to the outside. Though we use a
different formulation of the equations (a version of BSSN as opposed
to the Z4 system \cite{Bona04a}), this ``free black hole evolution''
approach is exactly the one that we analyze and apply in this paper. Even
though in several aspects this is different from the ``stuffed black hole
proposal'' \cite{Arbona97}, we will refer to our particular implementation as
{\it the relativistic turducken} \cite{Turducken}.

\section{No constraint leaking in the turduckening: an analytical proof}
\label{sec:formulation}

The exact version of the BSSN system we use is given by Eqs.\
(3)--(7), (21)--(23) in \cite{Beyer:2004sv} where we set the parameter
$m$ to one and the source terms $S$, $\hat{S}_{ij}$ and $S^i$ to
zero. Furthermore, the lapse $\alpha$ and the shift $\beta^i$ are
evolved according to the $1+\log$ slicing condition $\hat{\partial}_0
\alpha = -2\alpha K$ and the ``hyperbolic Gamma driver''
\cite{Alcubierre02a} like conditions $\hat{\partial}_0 \beta^i = 3
B^i/4$, $\hat{\partial}_0 B^i = \hat{\partial}_0\tilde{\Gamma}^i -
B^i/2$ , respectively, where $\hat{\partial}_0 = \partial_t -
\beta^j\partial_j$. The term $\hat{\partial}_0\tilde{\Gamma}^i$ in the
last equation is set equal to the right-hand side of the evolution
equations for the $\tilde{\Gamma}^i$ symbols. As noted in
\cite{Beyer:2004sv} the use of $\hat{\partial}_0$ (as opposed to
$\partial_t$) in the above equations simplifies the analysis of the
hyperbolic structure of the equations. Later, it was also found to be
important in practice for long-term binary evolutions
\cite{vanMeter:2006vi}. In addition, using $\hat{\partial}_0$ for the
lapse implies that the slicing obtained is independent of the choice
of shift vector \cite{Brown:2007}. 

The well-posedness of the resulting Cauchy problem was analyzed in
\cite{Beyer:2004sv}. A sufficient condition for well-posedness is
strong hyperbolicity of the evolution equations. (See
\cite{Kreiss:2001cu,Nagy:2004td} for definitions that apply to second
order systems.) In our case, the equations are strongly hyperbolic if
and only if the lapse $\alpha$ and the conformal factor $\phi$ are
smooth functions satisfying $\alpha > 0$, $|\phi| < \infty$ and
$h:=2\alpha-e^{4\phi}\ne 0$. The last condition is typically violated,
at least on some two-surface. This is so because in general, $\alpha
\to 1$ and $\phi \to 0$ and therefore $h\to 1$ as one approaches the
main asymptotically flat end, while near black holes $\alpha$ is small and
$\phi$ is large and positive (for the coordinate conditions used here
typically $\alpha\approx 0.3$ at the horizon, and $\alpha\to 0$ and
$\phi \to \infty$ at any punctures) so that $h < 0$ near a
horizon. Therefore, the function $h$ must be zero somewhere in
between. On the other hand, if the regions where $h = 0$ are, for
example, sets of zero-measure in the computational domain there is
hope that the violation of the condition $h \ne 0$ still allows for a
well posed Cauchy problem. The numerical simulations in Sect.\
\ref{sec:CP} below show no apparent sign of numerical instability.

The characteristic speeds (with respect to normal observers) for our
evolution equations are the following \cite{Beyer:2004sv}:
$0,\pm1,\pm\mu_1,\pm\mu_2,\pm\mu_3$, where $\mu_1 = \sqrt{2/\alpha}$,
$\mu_2 = \sqrt{3}\, e^{2\phi}/2\alpha$, $\mu_3 = e^{2\phi}/\alpha$.
It is possible to give a precise meaning to the different
characteristic fields and speeds in the high-frequency limit
\cite{Sarbach02b,Calabrese02c}. In that limit, fields propagating with
speeds $\mu_1$, $\mu_2$ and $\mu_3$ correspond to gauge modes, while
the fields corresponding to gravitational radiation and
constraint-violating modes have speeds $\pm 1$ and $0$, $\pm 1$ 
respectively. As we will see below, the constraint propagation system
possesses the characteristic speeds $0$ and $\pm 1$.

The BSSN system is subject to the Hamiltonian and momentum constraints
$H=0$ and $M_i=0$ plus three extra constraints associated with the
introduction of the $\tilde{\Gamma^i}$ symbol as independent
variables, namely $C^i_\Gamma := \tilde{\Gamma}^i +
\partial_j\tilde{\gamma}^{ij} = 0$, where $\tilde{\gamma}^{ij}$ refers
to the inverse of the conformal metric. In order to obtain a solution
to Einstein's vacuum field equations, these constraints have to be
satisfied. We now show that it is sufficient to solve them on an
initial Cauchy surface in the region exterior to the black holes. The
constraint propagation system then guarantees\footnote{However,
constraint violations can still be introduced by improper outer
boundary conditions.} that these constraints hold at every time future to
the initial surface and at every point outside the black hole regions,
\emph{independent of any constraint violation in the interior of the
black holes}. We show this by deriving the constraint propagation
system and casting it into first order symmetric hyperbolic
form. Then the causal propagation of the constraints can be shown via
a standard energy inequality provided all the characteristic speeds
(as measured by normal observers) of the system are smaller than or
equal to one in magnitude.

Using the Bianchi identities, imposing the evolution equations and
introducing the additional constraint variables $Z_{ij} = (\partial_i
C_\Gamma^k)\tilde{\gamma}_{kj}$, the constraint propagation system can
be rewritten as a first order system of the form
\begin{equation}
\hat{\partial}_0 C = \alpha\left[ 
{\bf A}(u)^i\partial_i C + {\bf B}(u) C \right],
\end{equation}
where $C$ are the constraint variables, $u =
(\alpha,\phi,\tilde{\gamma}_{ij},K,\tilde{A}_{ij},\tilde{\Gamma}^i)$
are the main variables, and ${\bf A}^1$, ${\bf A}^2$, ${\bf A}^3$ and
${\bf B}$ are matrix-valued functions of $u$. Decomposing $Z_{ij} =
\hat{Z}_{(ij)} + Z_{[ij]} + \gamma_{ij} Z/3$ into its trace-free
symmetric part $\hat{Z}_{(ij)}$, its anti-symmetric part $Z_{[ij]}$,
and its trace $Z = \gamma^{ij} Z_{ij}$ with respect to the physical
three-metric $\gamma_{ij}$, and representing $C$ in terms of the
variables $C = (C_\Gamma^i,S_1 := 2m H + Z, S_2 := H + 2\sigma Z, M_j,
\hat{Z}_{(ij)}, Z_{[ij]})$, the principal symbol ${\bf A}({\bf n}) =
{\bf A}(u)^i n_i$ is given by
\begin{eqnarray}
{\bf A}({\bf n}) C &=& \left( 0,0,n^j M_j\, ,
 \frac{1}{3} n_j S_2
 + \frac{1}{2} n^i \hat{Z}_{(ij)} + \frac{1}{2} n^i Z_{[ij]}\, , \right.
\nonumber\\
 && \left.\vphantom{\frac{1}{2}} \qquad
  2  (n_{(i} M_{j)})^{TF}, 2 n_{[i} M_{j]} \right),
\end{eqnarray}
where $n^i \equiv \gamma^{ij} n_j$ and $n_i$ is normalized such
that $n_i n^i = 1$. This system is symmetric hyperbolic, and its
characteristic speeds (with respect to normal observers) are  
$0$ and $\pm 1$. A symmetrizer is given by the quadratic form
\begin{eqnarray*}
C^T {\bf H} C &=& \tilde{\gamma}_{ij} C_\Gamma^i C_\Gamma^j + S_1^2
 + \frac{1}{3}\, S_2^2 + \gamma^{ij} M_i M_j\\
&& {} + \frac{1}{4}\,\gamma^{ik}\gamma^{jl}\hat{Z}_{(ij)}\hat{Z}_{(kl)}
 + \frac{1}{4}\,\gamma^{ik}\gamma^{jl} Z_{[ij]} Z_{[kl]}\; .
\end{eqnarray*}
The symmetrizer, along with the fact that there are no superluminal
characteristic speeds, allow us to obtain an energy estimate for the
constraint variables $C$ and to show that no constraint violations from
the interior of a black hole can propagate to the outside. The
explicit estimate will be presented elsewhere along with more details
of the results presented in this paper.

\section{Single black hole evolutions and the end state}
\label{sec:KS}

In this section we present insights obtained by applying 
the turducken technique to a single spherically symmetric black hole. 
For these studies we use both the three--dimensional (3D) code described in 
section IV, as well as the one--dimensional (1D) BSSN code discussed in 
\cite{Brown:2007a}. Both codes use a formulation 
of the BSSN equations that is strongly hyperbolic everywhere except in regions 
of the computational domain that are likely sets of measure zero, and have 
causal constraint propagation. 

For a single black hole we use turduckened Kerr--Schild (KS) initial data. Without turduckening, 
a KS slice hits the singularity. 
We first define the spacetime metric $g_{\mu\nu} = \eta_{\mu\nu} + 2H\ell_\mu \ell_\nu$ in terms of 
Cartesian coordinates $x$, $y$, $z$, where $\eta_{\mu\nu}$ is the Minkowski metric,
$H = M/\bar r$ and $\ell_\mu = (1,x,y,z)/\bar r$. Here, $\bar r$ is defined in terms of 
coordinate radius $r = (x^2 + y^2 + z^2)^{1/2}$ by 
$\bar r = (r^{p} + \epsilon^p)^{1/p}$. The contravariant metric $g^{\mu\nu}$ is obtained from 
$g_{\mu\nu}$ by raising indices with $\eta^{\mu\nu}$. In Cartesian coordinates the initial metric is defined by the 
spatial components of $g_{\mu\nu}$ and the initial extrinsic curvature is defined 
by the usual expression $K_{ij} = (-\dot g_{ij} + \beta^k \partial_k g_{ij} + 2g_{k(i} \partial_{j)}
\beta^k)/2\alpha$, where $\alpha = 1/\sqrt{-g^{tt}}$ and $\beta^i = -g^{it}/g^{tt}$. 
The initial data for the 1D code is obtained by transforming the Cartesian data to spherical coordinates. 

For $r \gg \epsilon$ we find $\bar r \approx r$ and the initial data coincides with a 
KS slice of a non--rotating black hole. 
For $r$ close to the origin, the data are smooth and regular as long as $\epsilon \ne 0$. 
This form of turduckening is not 
ideal since it leads to constraint violations that extend beyond the horizon $r=2\,M$. 
Typical values used in our simulations are $\epsilon = 0.1M$ and $p=4$. These values lead to 
initial violations of the Hamiltonian constraint of $\sim 10^4/M^2$ at $r=0$ and $\sim 10^{-5}/M^2$ 
at $r=2\,M$. 

Experiments in 1D show that after an evolution time of $50\,M$, the Hamiltonian constraint violation 
throughout the computational domain drops to a level $\sim 10^{-5}/M^2$. 
Similar results hold for the other constraints. By $t=50\,M$ 
the data have become nearly stationary; the final state in the $t\to\infty$ limit 
coincides with a portion of the stationary $1+\log$ slice of Schwarzschild. 
This is the same end state obtained with puncture evolution \cite{Hannam:2006vv,Brown:2007}. 
The key ingredient responsible for these remarkable behaviors is the Gamma--driver shift condition. With this condition the shift 
grows large in the interior region to counteract the grid stretching that would otherwise occur as the lapse 
collapses. As a result the time flow vector field tips outside the physical light cone (toward increasing $r$)
and the grid points near $r=0$ are quickly driven out of causal contact with the constraint violating 
portion of the initial data. 
With the constraints (nearly) satisfied everywhere in the computational domain, the numerical data represents
a slice of Schwarzschild that extends from region I of the Kruskal diagram, crosses the black hole horizon, and 
terminates at a resolution--dependent location inside the black hole. The $1+\log$ slicing condition then 
guides the slice to a stationary state. 

\begin{figure}
  \includegraphics[width=0.48\textwidth]{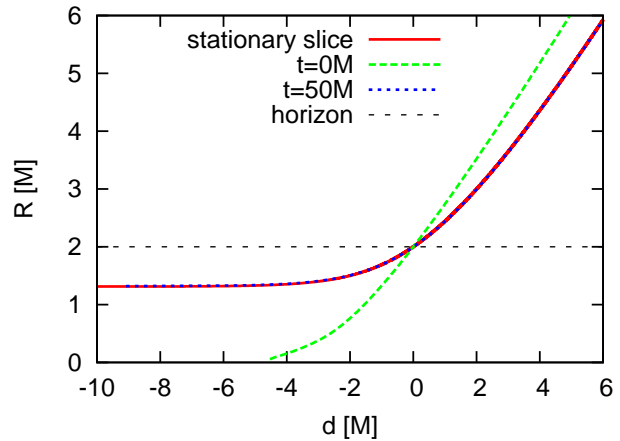}
  \caption{Areal radius $R$ versus proper distance $d$ from the
    horizon.  The initially turduckened KS become
    indistinguishable from the portion $d\, \greaterthansimilarto -9$ of a stationary $1+\log$ slice after
    $t\approx 50\,M$.  The region $R < 2\,M$ is the black hole interior.}
  \label{fig:rvp}
\end{figure}

Fig.\ \ref{fig:rvp} shows the areal radius $R$ versus proper
distance $d$ (in the radial direction) 
for a single non--spinning black hole, obtained from the 1D code with resolution $M/200$. 
 The data evolve to the stationary $1+\log$ slice in spite of the fact that the initial 
data violate the constraints. 

Alternatively, the initial KS data can be changed only inside a sphere of radius $r_0<2M$.
In 1D simulations such initial data can lead to the formation of gauge shocks, like those 
discussed in \cite{Garfinkle2007a}.  The shocks typically form just outside the 
black hole, independent of the parameter $r_0$; this suggests that the formation of shocks is a consequence of 
the gauge conditions, and not the black hole turduckening. 
We have not seen this behavior in 3D, perhaps due to lack of resolution. 

\section{Binary black hole evolutions using Cook-Pfeiffer data}
\label{sec:CP}

We evolve quasi-equilibrium binary black hole initial data using the
form of the equations described above, implemented in \texttt{CCATIE},
a three-dimensional adaptive mesh
refinement code which uses the Cactus framework \cite{cactusweb1_short} and
the Carpet mesh refinement driver
\cite{Schnetter-etal-03b, carpetweb_short}.  This evolution code is fourth
order accurate. It uses centered finite differencing
operators, except for the advection terms which are upwinded.
We use fifth order spatial and
second order temporal polynomial interpolation at mesh refinement boundaries,
and buffer zones as described in \cite{Schnetter-etal-03b} to ensure
stability.  We therefore
expect our code to be third order accurate in the limit of infinite
resolution, and expect it to show approximate fourth order convergence
away from the outer boundary and for the resolutions used here.
We use a fourth order Runge-Kutta time
integrator with a CFL factor of $0.4$.
We use Sommerfeld outer boundary conditions for the individual
components of the evolved variables, which
 are not constraint
preserving; we therefore place the outer boundaries at a large
distance from the source.

The initial data were provided by Harald Pfeiffer \cite{blackholesweb}
and are described in \cite{Cook:2004kt, Pfeiffer:2002wt}.  In particular
we use the data set \verb+sep_07.00_59a.tgz+ in which the binary
black hole system is expected to orbit approximately once before
merging.
These data have an ADM mass $M_{\mathrm{ADM}} \approx 2.44449 \approx
0.977795\,M$, where we use a scale factor $M=2.5$.  The black holes
are centered about $x = \pm1.4\,M$, and the apparent horizons
have a coordinate radius $r_{\mathrm{AH}}\approx 0.35\,M$.

Our simulations use reflection symmetry about $z=0$ and
$\pi$-rotation symmetry about the $z$ axis.  We choose a simulation domain
with outer boundaries at $204.8\,M$, and use altogether $9$
successively smaller levels of mesh refinement, where the finest level
has an extent of $0.8\,M$, centered about each black hole.  Our
resolution is $h=3.2\,M$ on the coarsest grid, $h=0.8\,M$ near the
gravitational wave detector, and $h=0.0125\,M$ on the finest grid.
We include results from two coarser runs with coarse grid resolutions
$h\approx4.5\,M$ and $h\approx4.1\,M$, respectively.

The initial data are provided in terms of spectral expansion
coefficients for 
the ADM variables on multiple domains and need to
be interpolated to our grid points.  The initial data setup excises
the apparent horizons but extrapolates a distance of up to 
$0.25\, r_{\mathrm{AH}}=0.0875\,M$ into the horizon.  The remainder of the
interior of the apparent horizons needs to be turduckened.

\begin{figure}
  \includegraphics[width=0.48\textwidth]{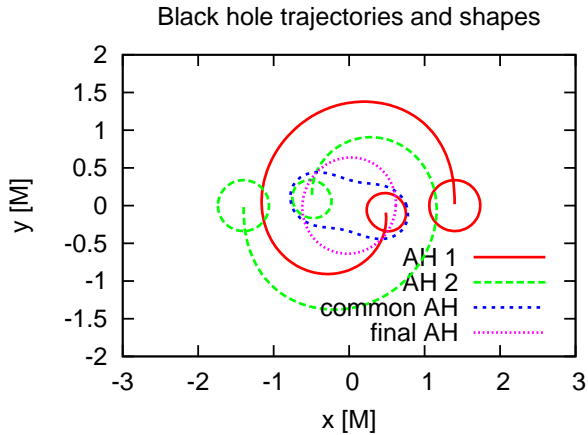}
  \caption{Apparent horizons in a binary black hole evolution of
    Cook-Pfeiffer data.  This figure shows the tracks of the centroids
    of the apparent horizons, as well as their shapes at $t=0\,M$, at
    merger ($t\approx37\,M$), and at late times ($t\ge200\,M$).  
  }
  \label{fig:cp-tracks}
\end{figure}


We have experimented with various methods for turduckening the black
hole interior, and find that the details do not matter much in
practice, as long as the spacetime remains unchanged within the finite
differencing stencil radius of the horizon.  Since there are
preciously few grid points between the excised region and the horizon,
we chose a method which leaves all given spacetime data unchanged and
fills in the excised points in a smooth manner.  (One alternative
would be a blending method which fills the excised region with
arbitrary data, and then modifies some of the non-excised grid points
to create a smooth match.)

In particular, we solve the elliptic equation $(\partial^6/\partial
x^6 + \partial^6/\partial y^6 + \partial^6/\partial z^6) A = 0$ to
fill the excised points of a quantity $A$, using standard centered
derivatives everywhere and using the given non-excised data as
boundary conditions where necessary.  This is equivalent to providing
boundary conditions for $A$ and its normal derivatives $\partial
A/\partial\mathbf{n}$, and $\partial^2 A/\partial\mathbf{n}^2$. The
result is therefore $C^2$ everywhere within the horizon. We solve this
equation with a standard conjugate gradient method.

\begin{figure}
  \includegraphics[width=0.48\textwidth]{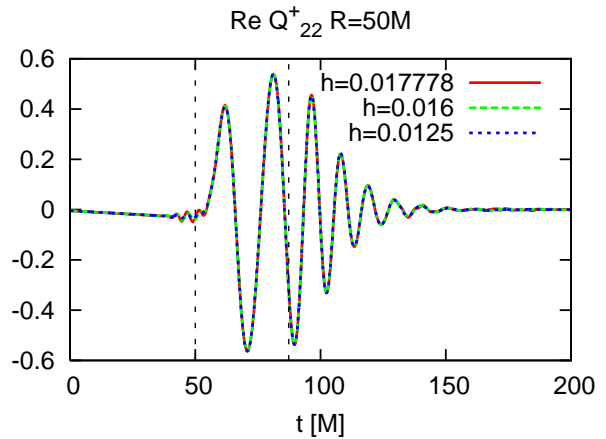}

  \caption{Real part of the waveform $Q^+_{22}$,
    extracted at $R=50\,M$.  The vertical lines
    indicate approximately when the initial burst of spurious
    radiation first reaches the detector and when the common horizon is
    ``seen'' by the detector.  The ``junk'' radiation near $t=50\,M$
    is a well-known feature from puncture evolutions.}
  \label{fig:cp-waveforma}
\end{figure}

\begin{figure}
  
  \includegraphics[width=0.48\textwidth]{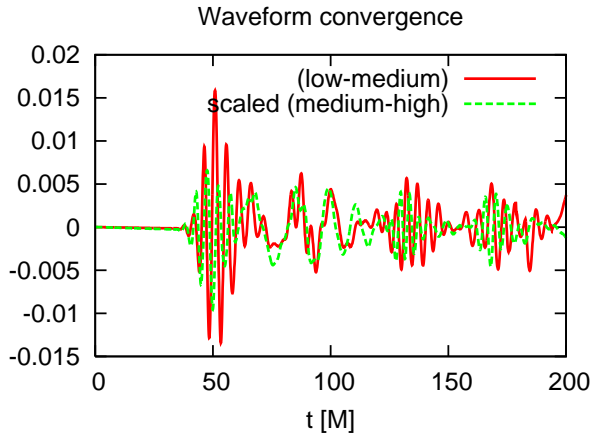}
  \caption{Difference between waveforms, scaled for 4th order
   convergence.  The waveform phases have been shifted in time so
   that all resolutions have the same phase at the beginning of the
   merger radiation burst at $t\approx60\,M$.  The initial
   high-frequency oscillations in the error are caused by the small
   amount of ``junk'' radiation near $t=50\,M$, and similar
   oscillations near $t=140\,M$ are probably caused by its reflection
   at the coarsest refinement level.  The noise near $t=170\,M$
   appears at a time when the waveform has already rung down.}
  \label{fig:cp-waveformb}
\end{figure}

We follow the evolution of these data through merger and ringdown for
about $200\,M$.  Fig.\ \ref{fig:cp-tracks} shows the locations,
shapes, and tracks of the individual and the common apparent horizons.  
A common horizon appears at about $t=37\,M$.  
The common horizon initially has a strong $Y_{22}$ deformation which is
radiated away. This is clearly shown in the real part of the
$\ell=m=2$ mode of the Zerilli function $Q^+$, extracted on a coordinate
sphere at $R=50\,M$ and shown in Fig.\ \ref{fig:cp-waveforma}. Fig.\
\ref{fig:cp-waveformb} shows the results of a convergence test,
although the resolutions are too close together to give reliable
results.
Both the horizon dynamics and waveforms are very similar to those from
puncture initial data.
We will present a study of this and other systems with larger
initial separations in more detail elsewhere.

\section{Final remarks}
\label{sec:Final}

A key property needed in a ``free black hole evolution'' approach is that the
constraints propagate causally. This cannot be taken for granted, and
must be proved (or tested) for any particular formulation of the
Einstein equations
used.  Note that  even apparently small
variations in the evolution system can change the constraint propagation 
from causal to acausal.

Causal propagation of the constraints alone is not sufficient. In
modifying the initial data
by smoothing away the singularity, we are not guaranteeing
that the evolution will proceed to a smooth, regular end-state.
That this end-state is numerically well-behaved
is the other key ingredient in any evolution
that relies on modifying the interior of the horizon in some way.
As the numerical evidence presented here shows,
evolutions in spherical symmetry do tend to a recognizable end-state for the
given set of gauge conditions and form of the equations.
It seems likely that a similar picture will hold away from
spherical symmetry.

Our work suggests that the turducken
technique will hold irrespective of how and when the data inside the horizon
are modified, thus allowing the method to be applied
without modification to the final stages of
evolutions performed with possibly different codes and/or methods, or
to horizons formed e.g.\ in stellar collapse scenarios.

\vspace{1ex}

Most of the results of this paper were originally presented
\cite{ecgm10web}
by one of us (ES) at the
Tenth Eastern Gravity Meeting. 
Since then, and while completing
this paper, independent results complementary to those presented here have been
presented in Ref.\ \cite{Etienne2007a}. 

\begin{acknowledgments}

We thank H. Pfeiffer for making the initial data available to us, and
for his help in implementing a reader for these data.
OS thanks D. Nu\~nez for help in deriving the characteristic speeds and
fields of the constraint propagation system. MT thanks S. Teukolsky for
hospitality at Cornell University, where part of this work was done. 
We employ J. Thornburg's apparent horizon finder
\cite{Thornburg2003:AH-finding}. This research was 
supported in part by NSF grant PHY 0505761 and NCSA grant MCA02N014 to
Louisiana State University, and by NSF grant PHY 0600402 to North 
Carolina State University.
We use the supercomputing resources Peyote at the AEI, Eric at LONI,
Supermike at LSU, and Abe and Tungsten at the NCSA\@.
We also employ the resources of the CCT at LSU, which is supported by funding from the Louisiana
legislature's Information Technology Initiative.  

\end{acknowledgments}

\bibliographystyle{bibtex/apsrev-nourl-fewauthors}
\bibliography{bibtex/references,localrefs}

\end{document}